\documentclass{svproc}
\usepackage{graphicx}

\usepackage{url}

\begin{document}
\mainmatter              
\title{Weak interference between the 1$^-$ states \\
in the vicinity of $\alpha$-particle threshold of $^{16}$O}
\titlerunning{Weak interference between the 1$^-$ states}  
%
\author{M. Katsuma\inst{1,2}}
\authorrunning{} 
%
\tocauthor{}
\institute{Advanced Mathematical Institute, Osaka City University, Japan,
  \and
Institut d'Astronomie et d'Astrophysique, Universit\'e Libre de Bruxelles, Belgium}

\maketitle              

\begin{abstract}
  The subthreshold 1$^-_1$ state at an excitation energy $E_x = 7.12$ MeV in $^{16}$O has been believed to enhance the $S$-factor of $^{12}$C($\alpha$,$\gamma$)$^{16}$O.
  The enhancement seems to originate from strong interference between 1$^-_1$ and 1$^-_2$ ($E_x\approx 9.6$ MeV) in the vicinity of the $\alpha$-particle threshold.
  However, weak interference between them and a resulting small $E$1 $S$-factor are exemplified with $R$-matrix theory.
  Including a higher-order correction of the resonance parameters, the present example appears to reproduce the experimental data consistently.
  It would therefore be possible that the $E$1 $S$-factor is reduced at low energies.
  \keywords{$^{12}$C($\alpha$,$\gamma$)$^{16}$O, $R$-matrix method, Stellar nucleosynthesis}
\end{abstract}

  The 1$^-_1$ ($E_x= 7.12$ MeV) and 1$^-_2$($E_x\approx 9.6$ MeV) states in $^{16}$O play an important role in the low-energy extrapolation of $^{12}$C($\alpha$,$\gamma$)$^{16}$O cross sections.
  If complicated process of compound nuclei is assumed, strong interference between them is expected, and $E$1 transition becomes predominant.
  At present, this interference has been believed to describe the cross section at $E_{c.m.}= 300$ keV.
  However, I have predicted a small $E$1 $S$-factor at this energy from the potential model (PM) \cite{Kat08}, because non-absorptive scattering results in weak coupling between shell and cluster structure in $^{16}$O.
  Besides, I have shown that $E$2 transition is dominant because 2$^+_1$ ($E_x= 6.92$ MeV) has $\alpha$+$^{12}$C structure \cite{Kat10,Fuj80}.

  In this paper, weak interference between 1$^-_1$ and 1$^-_2$, and the resulting small $E$1 $S$-factor are exemplified with $R$-matrix theory \cite{Kat17}.
  I estimate their reduced $\alpha$-particle widths from \cite{Kat08,Kat10}, and use the conventional $R$-matrix method \cite{Des03,Tho09}.
  In addition, the formal parameters are obtained from an exact expression, including a higher-order correction, because it has been reported that the parameters for 1$^-_2$ are not appropriately treated in the linear approximation \cite{Des03}.
  This correction ensures that $R$-matrix calculations correspond to the experimental data.

  Before showing an example of calculations, let me describe the $R$-matrix parameters.
  The Schr\"odinger equation is solved with the $R$-matrix,
  \begin{eqnarray}
    R_L(E_{c.m.}) &=&
    \mathop{\sum}_{n} \frac{\tilde{\gamma}_{n L}^2}{\tilde{E}_{n L}-E_{c.m.}} + {\cal R}_{\alpha L},
    \label{eq:a}
  \end{eqnarray}
  where ${\cal R}_{\alpha L}$ is the non-resonant component.
  $\tilde{E}_{nL}$ and $\tilde{\gamma}_{nL}$ are the {\it formal} resonance energy and {\it formal} reduced width, respectively.
  These are different from the Breit-Wigner (observed) parameters, $E_{nL}$, $\gamma_{nL}$.
  The conversion is given as

  \begin{eqnarray}
    \tilde{E}_{n L}(E_{c.m.}) &=&
    E_{n L} + \tilde{\gamma}_{n L}^2(E_{c.m.})\Delta_L(E_{n L}, a_c) \,[\,1+d_{n L}\,]
    \label{eq:b}
    \\
    \tilde{\gamma}_{n L}^2(E_{c.m.}) &=&
    \frac{\gamma_{n L}^2}{1-\gamma_{n L}^2\Delta_L^\prime(E_{n L},a_c)\left[\,1+Q_{n L}(E_{c.m.},a_c)\,\right]},
    \label{eq:c}
  \end{eqnarray}
  where $Q_{nL}$ is the higher-order correction of the resonance parameters, depending on energies.
  Note that $Q_{nL}=0$ was used in most of reactions \cite{Des03,Tho09}.
  $\Delta_L$ is the shift function, $\Delta_L^\prime=d\Delta_L/dE$.
  $d_{n L}$ is a parameter for multi-levels, and it is adjusted self-consistently so as to satisfy $\Delta_L(E_{nL},a_c) R_L(E_{nL}) = 1$; $d_{11}=-1.0133$, $d_{21}=0$.
  The observed parameters are 1$^-_1$: $E_{11}=-0.0451$ MeV, $\gamma_{11}=0.345$ MeV$^{1/2}$;
                              1$^-_2$: $E_{21}= 2.434$ MeV,  $\gamma_{21}=0.850$ MeV$^{1/2}$.
  $a_c$ is the channel radius, $a_c=4.75$ fm.
  All nucleons are interacting close together in the internal region,
  whereas nucleons are well-separated into $\alpha$ and $^{12}$C outside the region.
  Other observed parameters are taken from \cite{Til93}.
  ANC of 1$^-_1$ is $5.0\times10^{28}$ fm$^{-1}$ \cite{Kat08,ANC}.

\begin{figure}[t]
  \begin{center}
    \begin{tabular}{cc}
      \includegraphics[width=0.375\linewidth]{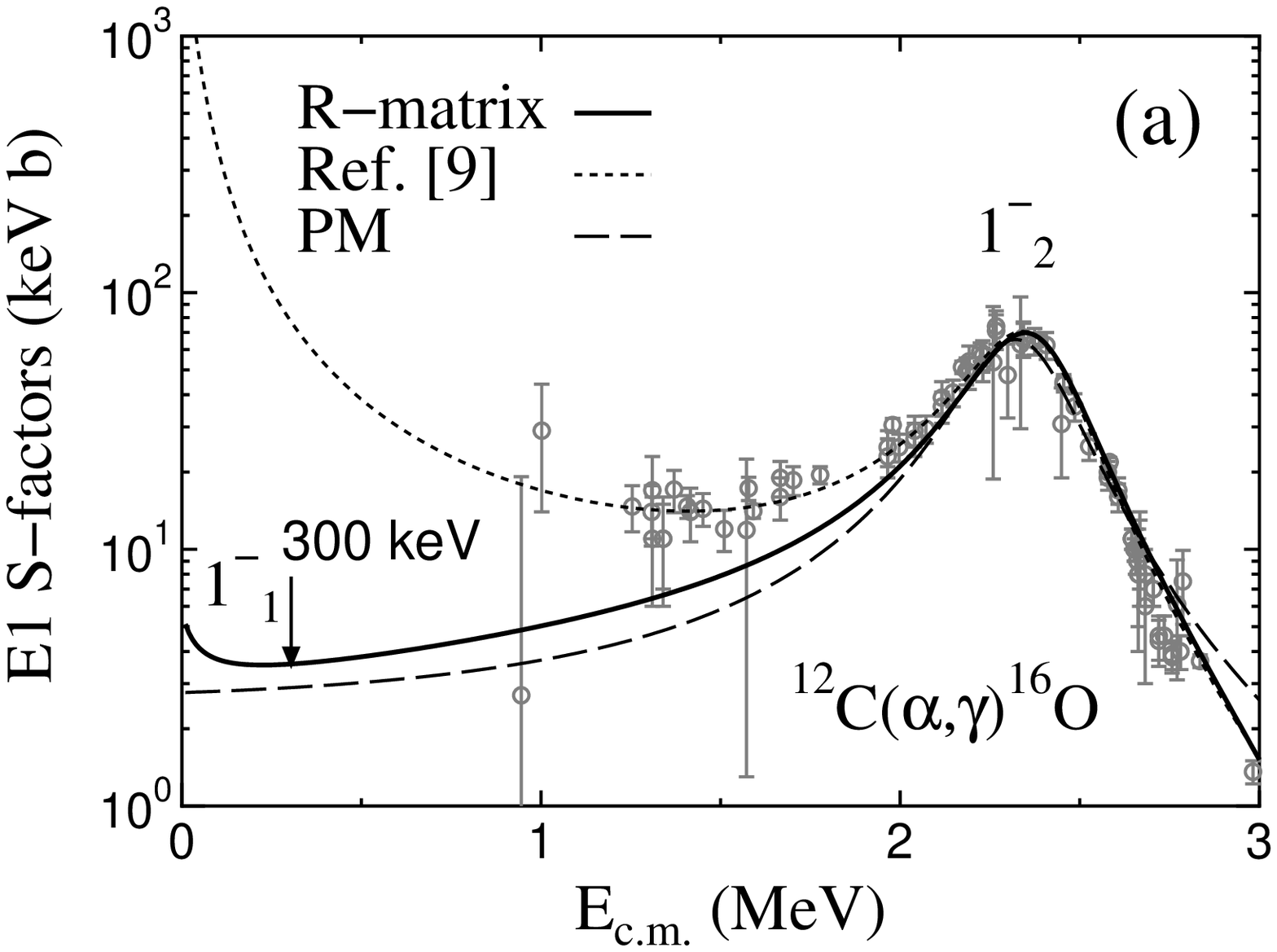} &
      \includegraphics[width=0.375\linewidth]{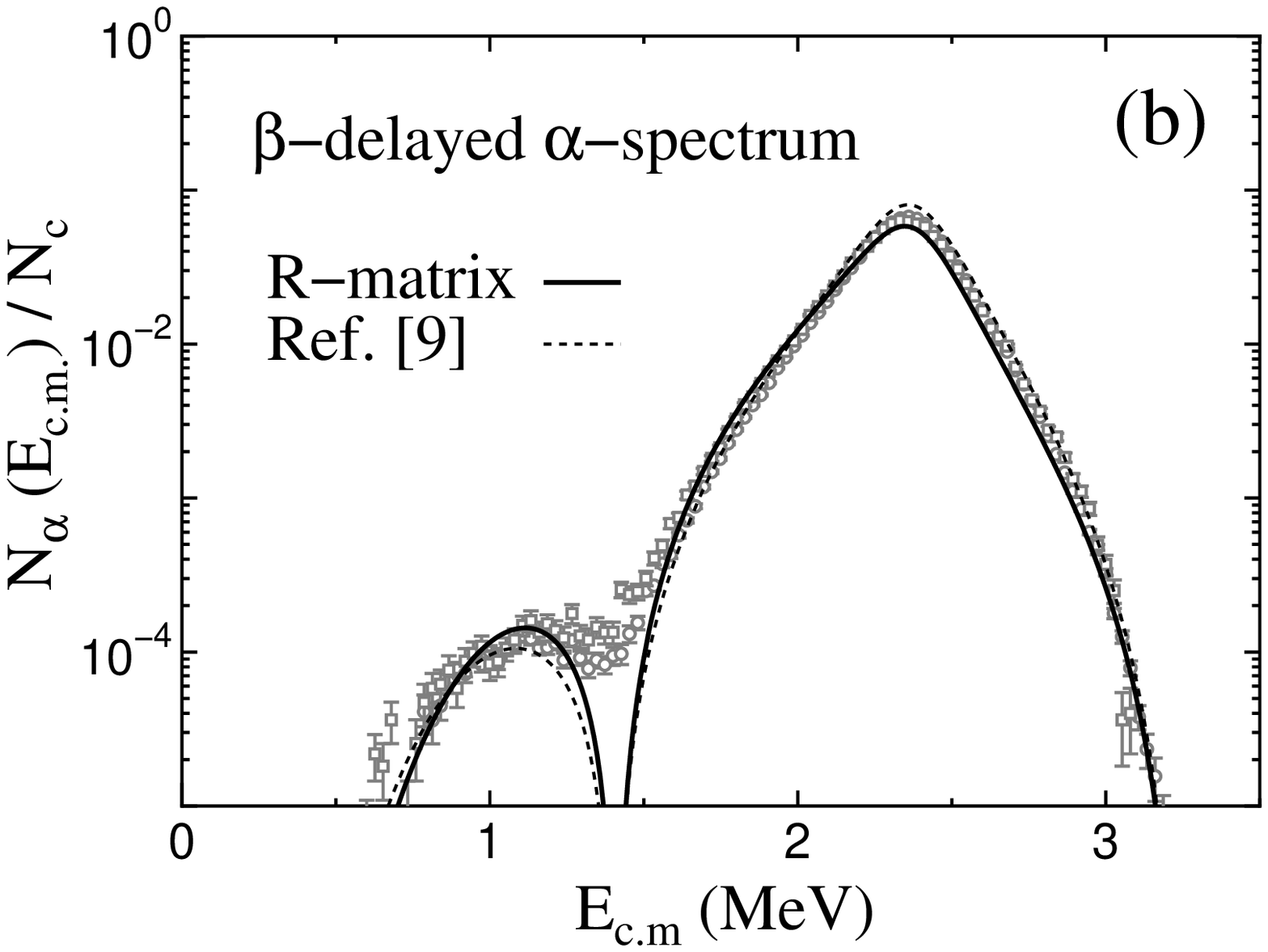}\\
      \includegraphics[width=0.375\linewidth]{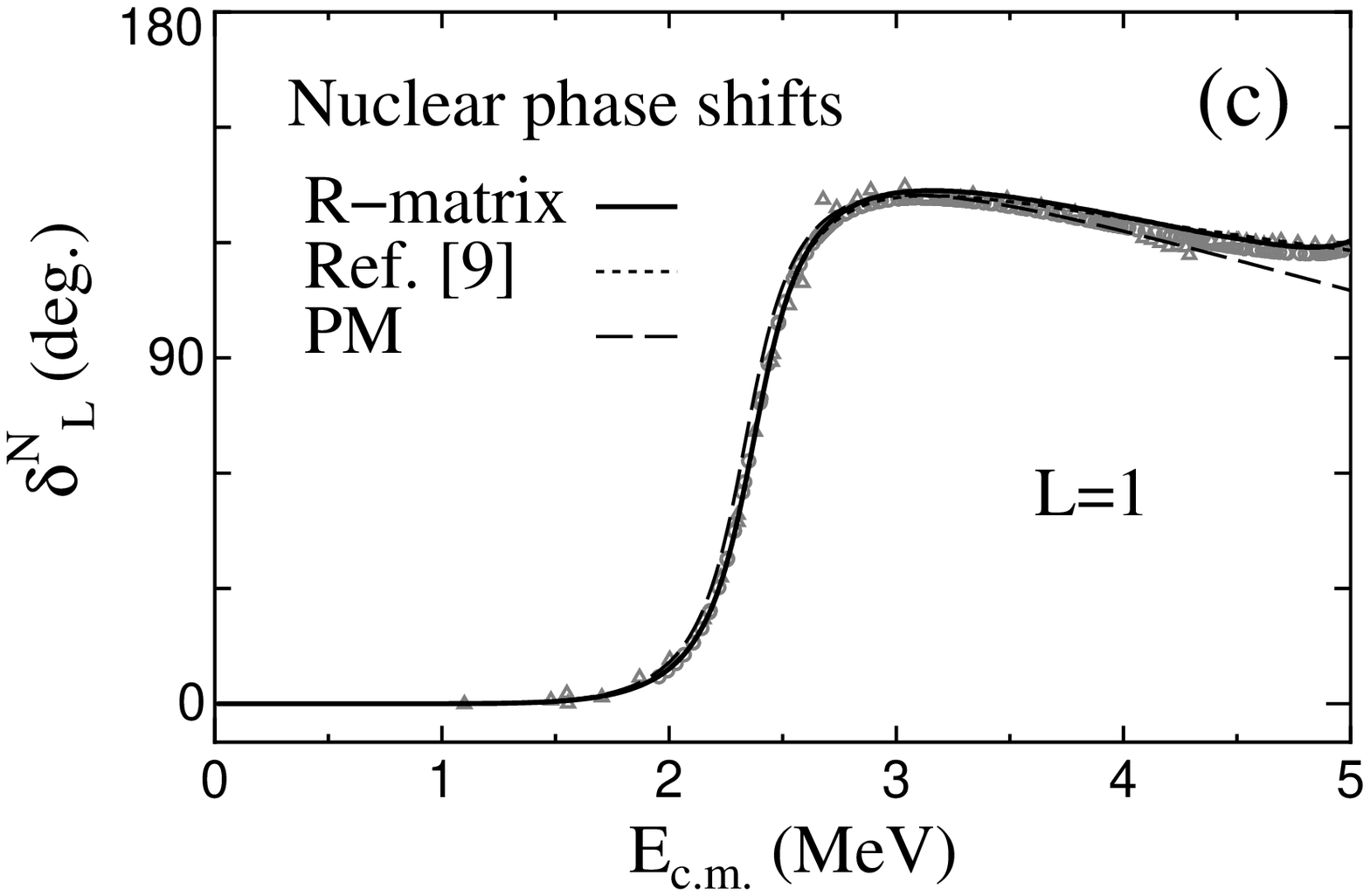} &
      \includegraphics[width=0.375\linewidth]{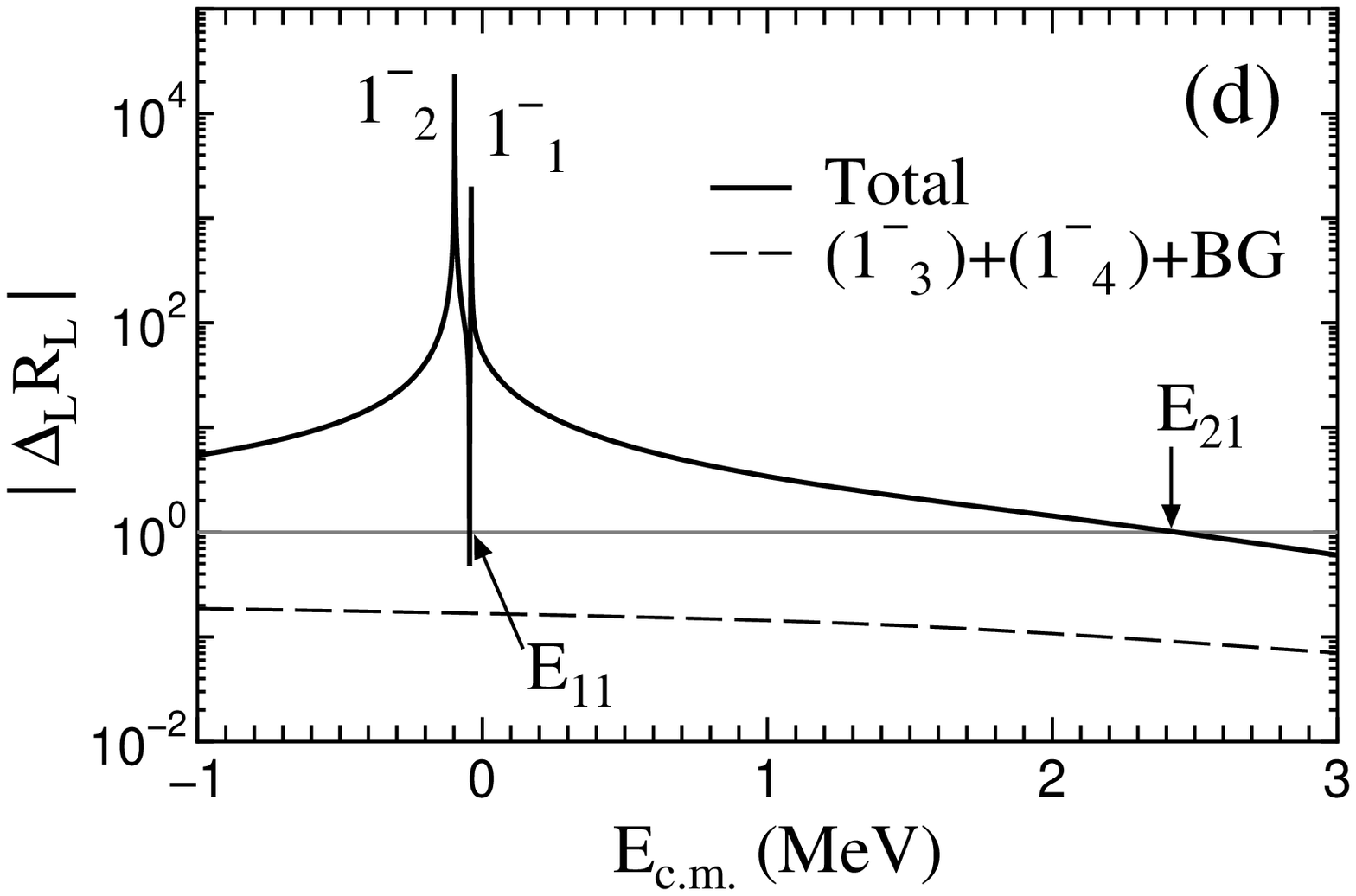}\\
    \end{tabular}
    \vspace{-6mm}
  \end{center}
  \caption{\label{fig:1} An example of $R$-matrix calculations of (a) $E$1 $S$-factor of $^{12}$C($\alpha$,$\gamma_0$)$^{16}$O, (b) $\beta$-delayed $\alpha$-particle spectrum of $^{16}$N, (c) $p$-wave phase shift of $\alpha$+$^{12}$C elastic scattering.
    The solid, dotted, and dashed curves are the results of the present work, $R$-matrix method \cite{Azu94}, and PM \cite{Kat08}, respectively.
    The experimental data are taken from \cite{Azu94,Tan10,exps}.
    (d) The resultant $R$-matrix is illustrated with the solid curve.
    The dashed curve is the sum of 1$^-_3$, 1$^-_4$, and non-resonant components.
  }
\end{figure}

  The example of the small $E$1 $S$-factor is shown by the solid curve in Fig.~\ref{fig:1}(a).
  The present example includes the component of the subthreshold state, and it resembles PM \cite{Kat08} (dashed curve).
  The interference between 1$^-_1$ and 1$^-_2$ appears to be weak.
  The corresponding calculations of the $\beta$-delayed $\alpha$-particle spectrum of $^{16}$N and the $p$-wave phase shift of $\alpha$+$^{12}$C elastic scattering are consistent with the experimental results \cite{Azu94,Tan10}. (Fig.~\ref{fig:1}(b) and \ref{fig:1}(c))
  So, the small $E$1 $S$-factor in Fig.~\ref{fig:1}(a) is in agreement with these experimental data.
  The experimental $\alpha$-particle width of 1$^-_2$ ($\Gamma_\alpha^{\rm exp}= 420\pm 20$ keV \cite{Til93}) is also reproduced by the present example, $\Gamma_\alpha^{\rm th}= 432$ keV.
  The dotted curves are the $R$-matrix calculations \cite{Azu94} with $Q_{nL}=0$, in which the narrow reduced widths are assumed.
  The derived 1$^-_2$ width \cite{Azu94} does not reproduce the experimental one.
  Compared with the solid curves, $Q_{nL}$ is found to reduce the $E$1 $S$-factor at low energies.
  In fact, a large energy shift for 1$^-_2$ is expected from the large reduced width of $\alpha$+$^{12}$C cluster structure. (Eq.~(\ref{eq:b}))
  So, the resultant energy of the 1$^-_2$ pole is found to be located in the vicinity of 1$^-_1$. (Fig.~\ref{fig:1}(d))
  This proximity of the poles suppresses their interference, and it consequently makes the small $E$1 $S$-factor below the barrier.

  The present example can be replaced with my previous result from PM, so I could use a hybrid model \cite{Lan83}, $E$1($R$-matrix)+$E$2(PM).
  The resulting total $S$-factor and reaction rates are confirmed to be concordant with \cite{Kat08,Kat12}.

  In summary, the weak interference between 1$^-_1$ and 1$^-_2$, and the small $E$1 $S$-factor have been exemplified with $R$-matrix theory.
  The formal parameters are obtained from the exact expression, including the higher-order correction.
  The reduced $\alpha$-particle widths of 1$^-_1$ and 1$^-_2$ are estimated from PM.
  The present example is consistent with the experimental results of the $\beta$-delayed $\alpha$-spectrum of $^{16}$N, the $p$-wave phase shift, and the $\alpha$-decay width of 1$^-_2$.
  In the example, the pole energy of 1$^-_2$ is located in the vicinity of 1$^-_1$.
  This proximity suppresses their interference, and it makes the small $E$1 $S$-factor below the barrier.
  It would therefore be possible in the $R$-matrix method that the $E$1 $S$-factor is reduced from the enhanced value currently expected.
  At the same time, the reaction rates are confirmed to be obtained from the direct-capture mechanism \cite{Kat08,Kat12}.

  I am grateful to Prof. S.~Kubono for his comments.
  I also thank M.~Arnould, A.~Jorissen, K.~Takahashi, H.~Utsunomiya, Y.~Ohnita, and Y.~Sakuragi for their hospitality at Universit\'e Libre de Bruxelles and Osaka City University.

%

\end{document}